\documentclass{iaus}
\usepackage{graphicx}

\newcommand{\arcm}{\mbox{\ensuremath{^{\prime}}}}

\title[The Massive Star Forming Region, Cygnus OB2]
{The Massive Star Forming Region, \\ Cygnus OB2}

\author[N.J. Wright]
{N.J. Wright and J.J. Drake}

\affiliation{Harvard-Smithsonian Center for Astrophysics, 60 Garden Street, Cambridge, MA 02138, USA}

\pubyear{2009}
\volume{266}
\pagerange{119--126}
\jname{Star Clusters: Basic Galactic Building Blocks Throughout Time And Space}
\editors{Richard de Grijs and Jacques Lepine}
\begin{document}

\maketitle

\begin{abstract}

We present results from a catalogue of 1696 X-ray point sources detected in the massive star forming region (SFR) Cygnus~OB2, the majority of which have optical or near-IR associations. We derive ages of 3.5 and 5.25 Myrs for the stellar populations in our two fields, in agreement with recent studies that suggest that the central 1-3 Myr OB association is surrounded and contaminated by an older population with an age of 5-10 Myrs. The fraction of sources with proto-planetary disks, as traced by K-band excesses, are unusually low. Though this has previously been interpreted as due to the influence of the large number of OB stars in Cyg~OB2, contamination from an older population of stars in the region could also be responsible. An initial mass function is derived and found to have a slope of $\Gamma = -1.27$, in agreement with the canonical value. Finally we introduce the recently approved Chandra Cygnus OB2 Legacy Survey that will image a 1 square degree area of the Cygnus~OB2 association to a depth of 120~ks, likely detecting $\sim$10,000 stellar X-ray sources.


\end{abstract}

\firstsection

\section{Introduction}

Massive star forming regions (SFRs) are thought to be the major star factories in the Universe (Elmegreen 1985), each containing hundreds to thousands of OB stars and millions of low mass stars. Understanding how these regions form and evolve is vital to our comprehension of the factors controlling the stellar initial mass function and the structure and evolution of our Galaxy. Incisive studies of massive SFRs are hindered by their scarcity and great distances, resulting in both an inability to probe the low mass end of the stellar spectrum and insufficient spatial resolution to diagnose the physical processes at work.
Therefore it is important to make full use of the very few nearby opportunities we have to pursue these issues. Cygnus OB2 is one such region at a distance of only 1.45~kpc (Hanson 2003) and is one of the most massive OB associations in our Galaxy with a rich and diverse stellar population (e.g. Massey \& Thompson 1991).
X-ray observations are highly effective in separating young association members from older field stars because pre-main-sequence (pre-MS) stars are thousands of times more luminous in X-rays than MS stars (Preibisch \& Feigelson 2005). To penetrate the high extinction in the direction of Cygnus OB2 as well as removing the contamination from Galactic field stars, we have used Chandra observations of Cyg OB2 to study this massive star forming region on our doorstep. 

\section{Observations}

We have combined X-ray observations of two Chandra fields in the center of Cygnus~OB2 (Figure~1) with optical and near-IR data from the INT Photometric H$\alpha$ Survey (IPHAS, Drew et al. 2005), the 2~Micron All Sky Survey (Cutri et al. 2003), and the United Kingdom Infrared Deep Sky Survey (Lawrence et al. 2007) surveys. Our catalogue (Wright \& Drake 2009) contains 1696 X-ray sources with broad-band fluxes, spectral fits, and variability information, of which 1501 (89\%) have optical or near-IR counterparts. From an analysis of the Poisson false-source probabilities we estimate that this catalogue contains $< 1$\% of false sources, and an even lower fraction when only sources with optical or near-IR associations are considered. A Monte Carlo simulation of the Bayesian cross-matching scheme also allowed the various sources of error in this process to be quantified. IPHAS photometry was employed to remove contamination from foreground stars, $\sim$5\% of the catalogue. Simulations of the extragalactic X-ray population were used to estimate its impact on the catalogue. The majority of extragalactic sources are not expected to have optical or near-IR counterparts due to the high Galactic extinction in Cygnus.

\begin{figure}[t]
\begin{center}
 \includegraphics[width=5.2in]{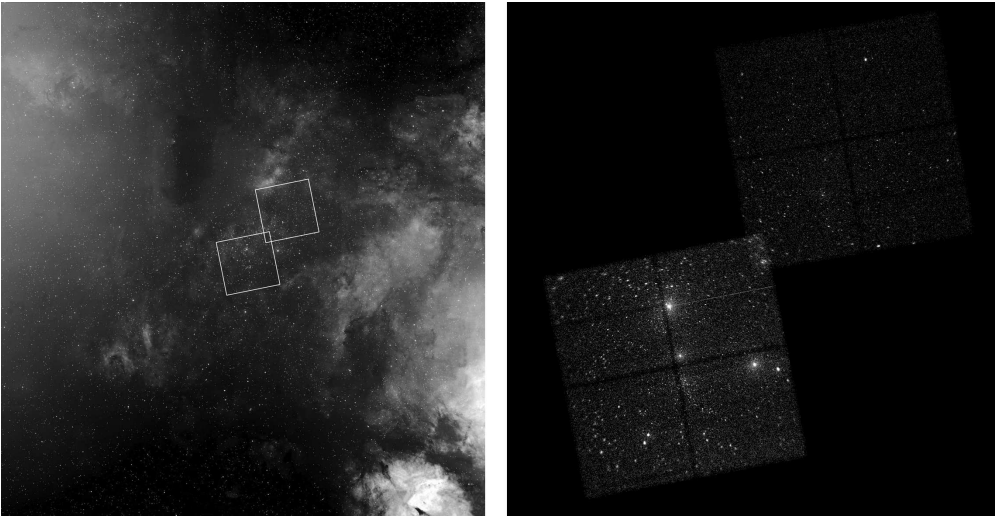} 
 \caption{{\it Left:} H$\alpha$ image of the Cygnus OB2 region mosaiced from IPHAS observations and displayed using a logarithmic intensity scale. The image size is $2.67 \times 2.85$ degrees and centered on (RA, decl.) = (20:33:00, +41:15:00) with north up and east to the left. The $17\arcm \times 17\arcm$ fields of view of the two {\it Chandra}-ACIS observations discussed in this paper are shown as white squares.
{\it Right:} Greyscale image (approximately 30\arcm~$\times$~30\arcm) of the two Chandra ACIS-I observations of Cygnus OB2. Image intensity is proportional to the log of the X-ray count density.
}
\end{center}
\end{figure}

\section{Near-IR Stellar Properties}

The near-IR color-color and color-magnitude (Figure~2) diagrams were used to study the stellar populations in the two Chandra fields. Pre-MS stellar isochrones (Siess et al. 2000) were used to estimate mean ages of 3.5 and 5.25 Myrs for the two fields. Considerable spread was observed around the isochrones, despite the low photometric errors, potentially suggesting a large age spread in the region. These results are in agreement with recent studies (e.g. Drew et al. 2008, Comeron et al. 2008) that suggest that the central association of OB stars with an age of 1-3 Myrs is surrounded and contaminated by an older population with an age of  $\sim$5-10 Myrs.
Proto-planetary disks are traced using the near-IR K-band excess, though only 5.8\% (central field) and 7.6\% (NW field) of sources are found to have inner disks. These low disk fractions could be attributed to the disrupting effects of the strong stellar winds and UV radiation from the OB stars in the vicinity, or they may be due to the influence of older stellar populations in the region.

\begin{figure}[t]
\begin{center}
 \includegraphics[width=3.0in, angle=270]{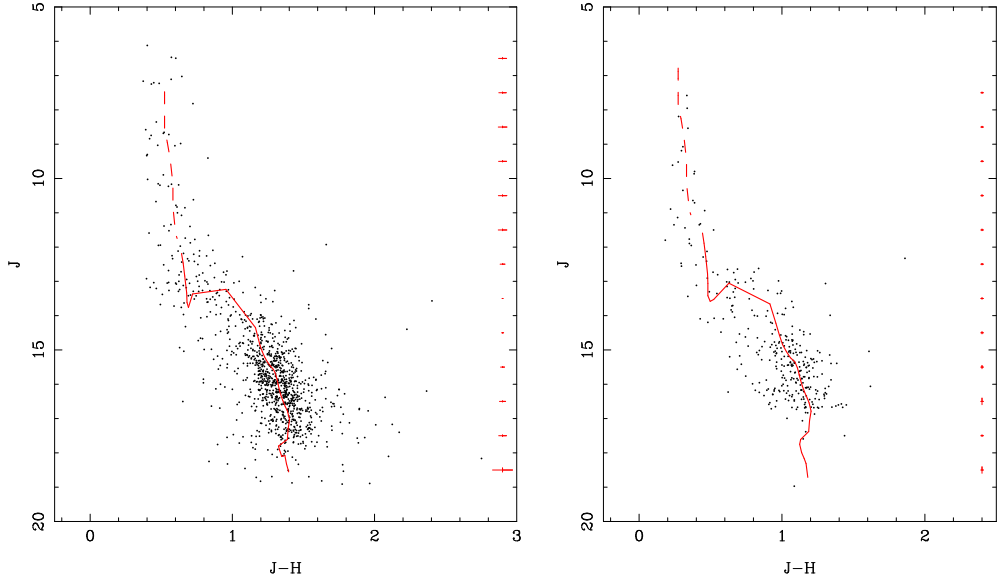} 
 \caption{Near-IR color-magnitude diagrams for sources in the central (left) and north-western (right) fields. Zero age MS (dashed red line) and pre-MS (full red line) tracks are shown. Mean photometric errors per magnitude bin are shown on the right of each figure.}
\end{center}
\end{figure}

\section{X-ray Stellar Properties}

The X-ray luminosity function (XLF) of Chandra sources was compared to that of the well studied Orion Nebula Cluster (Getman et al. 2005). We find a good agreement between the two, suggesting we are not missing any significant component. Within the limits of X-ray observations (i.e. that late B and early A-type stars do not all emit X-rays), we find that we are complete to 1 M$_{\odot}$ in both fields. The distribution of median X-ray source energies provides no evidence for a population of hard X-ray sources that would originate from highly embedded sources. This lack of embedded sources suggests that star formation in these two fields has ceased.

\section{The Initial Mass Function}

Using masses derived from existing spectroscopy and fits to the near-IR CMD we derived stellar masses for all our sources. The resulting mass function (Figure~3), complete to 1 M$_{\odot}$, has a slope of $\Gamma = -1.27 \pm 0.16$, in good agreement with the canonical value (Kroupa 2002). There is no evidence for a shallow mass function or a high-mass turnover in this mass range as has been predicted for some massive star forming regions. There is only a small difference in slope between the two fields (Figure~3), suggesting that significant mass segregation has yet to occur despite the age of the region. The mass function is steeper at higher masses with $\Gamma = -2.19 \pm 0.07$, which could indicate a steepening of the mass function, or could be due to contamination of a population of older stars in the region that have lost their highest-mass members.

\begin{figure}[t]
\begin{center}
 \includegraphics[width=2.05in, angle=270]{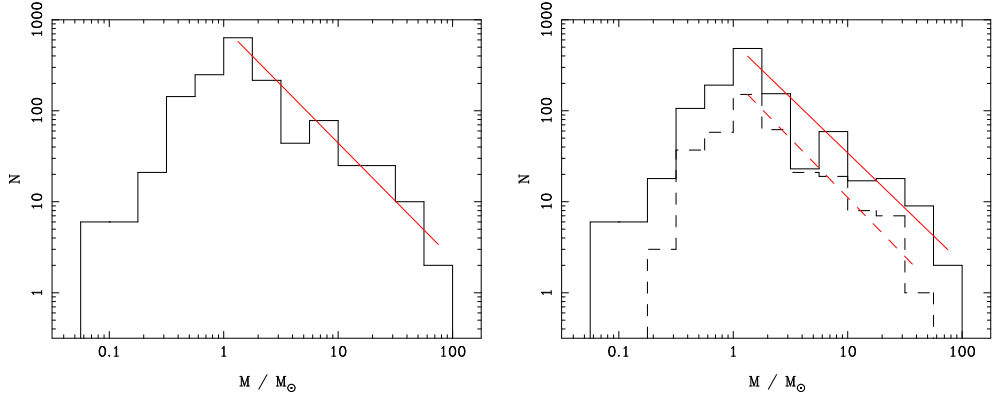} 
 \caption{Mass functions for sources in both fields (left) and in the separate fields (right; central field - full line, NW field - dashed line). Power-law fits are shown in red excluding the bins containing A and B-type stars. The power law indices are $\Gamma = -1.27 \pm 0.16$ (combined fields), $\Gamma = -1.22 \pm 0.16$ (central field) and $\Gamma = -1.29 \pm 0.16$ (NW field).}
\end{center}
\end{figure}

\section{The Chandra Cygnus OB2 Legacy Survey}

These observations cover only a very small region in the centre of Cygnus OB2, which Knodlseder (2000) suggest may be as large as 2 degrees wide. The recently approved Chandra Legacy Survey of Cygnus OB2 (PI: J.J. Drake) will extend these observations to a 1 degree wide region (Figure 5) with a depth of 120ks, achieving a completeness down to 0.5 M$_{\odot}$ and detecting $\sim$10,000 cluster members.
These observations, combined with existing (e.g. VLA, IPHAS, UKIDSS, Spitzer) and approved (e-Merlin, VERITAS, Herschel) surveys of the region, will provide a huge dataset for deep studies of star and planet formation in a massive star forming region.

\section{Conclusions}

First results are presented from the recently published catalogue of X-ray sources in the massive star forming region Cygnus~OB2 (Wright \& Drake 2009). The catalogue contains 1696 X-ray point sources, of which 1501 (89\%) have optical or near-IR counterparts. The XLF of sources indicates that, within the limits of an X-ray selected sample, we are complete to $\sim$1~M$_{\odot}$. The near-IR color-magnitude diagram suggests a potential age spread across the region, and best fitting ages of 3.5 and 5.25~Myr support recent suggestions for multiple epochs of star formation in Cygnus~OB2. The mass function is measured and found to have a slope of $\Gamma = -1.27$, in agreement with the canonical value and providing no evidence for a shallower slope, as predicted for some massive SFRs.

\end{document}